\documentclass[doublecol]{epl2} 
\usepackage{amssymb}
\usepackage{color}

\title{A Comparative Test of the $\Lambda$CDM and $R_{\rm h}=ct$ Cosmologies Based on Upcoming 
Redshift Drift Measurements}
\shorttitle{Model Selection with Redshift Drift Measurements} 

\author{F. Melia
}
\shortauthor{F. Melia}

\institute{                    
Department of Physics, The Applied Math Program, and Department of Astronomy, \\
The University of Arizona, AZ 85721, USA}
\pacs{98.80.-k}{Cosmology}
\pacs{98.80.Jk}{Relativistic Astrophysics}

\abstract{A measurement of the redshift drift constitutes a model-independent probe of fundamental 
cosmology. Several approaches are being considered to make the necessary observations, using
(i) the Extremely Large Telescope (ELT), (ii) the Cosmic Accelerometer, and (iii) the 
differential redshift drift methodology. Our focus in this {\it Letter} is to assess how these 
upcoming measurements may be used to compare the predictions of $\Lambda$CDM with those 
of the alternative Friedmann-Lema\^itre-Robertson-Walker cosmology known as the $R_{\rm h}=ct$ 
universe, and several other models, including modified gravity scenarios. The ELT should be 
able to distinguish between $R_{\rm h}=ct$ and the other models at better than $3\sigma$ 
for $z\gtrsim 3.6$ after 20 years of monitoring, while the Cosmic Accelerometer may be able 
to achieve the same result with sources at $z\gtrsim 2.6$ after only 10 years.}

\begin{document}

\maketitle

\section{Introduction}\label{intro}
Ongoing comparative studies between the {\it Planck} $\Lambda$CDM
(standard) model \cite{Ostriker:1995,Planck:2016} and the alternative
Friedmann-Lema\^itre-Roberson-Walker cosmology known as the
$R_{\rm h}=ct$ universe \cite{Melia:2007,MeliaShevchuk:2012,Melia:2020},
suggest that the latter may be a better fit to the data at both high and
low redshifts \cite{Melia:2022e}. For example, several anomalies with the
distribution of anisotropies studied in the cosmic microwave background
with $\Lambda$CDM as the background cosmology \cite{Bennett:2013,Planck:2016a}
appear to be completely mitigated by $R_{\rm h}=ct$
\cite{MeliaLopez:2018,Melia:2021b,Sanchis-Lozano:2022}.

In addition, while the baryon acoustic oscillations measured in the Ly-$\alpha$
forest with the final SDSS-IV release \cite{Blanton:2017} appear to disagree
with the standard prediction at a confidence level of $\sim 2.5\sigma$, they
favor $R_{\rm h}=ct$ over $\Lambda$CDM with a likelihood of $\sim 80\%$ versus
$\sim 20\%$ \cite{Melia:2023d}. Most recently, we have also witnessed a growing
discordance between the expected formation of structure in $\Lambda$CDM and
the discovery by {\it JWST} of well-formed galaxies at $z\gtrsim 16$
\cite{Pontoppidan:2022,Finkelstein:2022,Treu:2022} and an X-ray bright
supermassive quasar at $z\sim 10.1$ \cite{Bogdan:2023}. In both cases,
the time versus redshift relation in the standard model does not work
at all with the premature emergence of such objects so soon after the
big bang, but the required timeline is fully consistent with the predictions
of $R_{\rm  h}=ct$ \cite{MeliaMcClintock:2015,Melia:2023}. By now, this
contrast follows a well formed pattern that also accounts for the
discovery of $\sim 2$ Glyr-sized structures, which are ten times larger
than can be accommodated in $\Lambda$CDM, yet well within the
expected evolution in the $R_{\rm h}=ct$ cosmology \cite{Melia:2023c}.

A short summary of the fundamental basis for $R_{\rm h}=ct$,
and its comparative tests with $\Lambda$CDM based on many kinds of 
cosmological data, is provided in the Appendix.
All in all, the evidence favoring $R_{\rm h}=ct$ over $\Lambda$CDM appears
to be compelling (see also Table~2 in ref.~\cite{Melia:2018e}), but
the most definitive test ruling out one or the other of these two
competing models is arguably the measurement of redshift drift. Unlike
other kinds of cosmological probes, the redshift drift of sources
moving passively with the Hubble flow allows us to see the Universe
expand in real time \cite{Sandage:1962,McVittie:1962,Loeb:1998}.
Crucially, the standard model predicts a non-zero drift, while
$R_{\rm h}=ct$ predicts absolutely zero drift at all redshifts.
The distinction could not be clearer. For these two models, the
test produces an essentially yes/no outcome, regardless of what
the actual drift turns out to be if it's not zero.

The first detailed study to gauge the observational feasibility of
such a measurement was carried out by \cite{Liske:2008}, and this
has become a key science driver for the ELT-HIRES spectrograph \cite{Liske:2014}.
Since then, two other approaches have been proposed. In the first
of these, the Cosmic Accelerometer \cite{Eikenberry:2019} is focused
on lowering the cost of the project. The second, known as The
Acceleration Programme \cite{Cooke:2020}, aims to measure the
differential redshift drift between pairs of non-zero redshifts
to maximize the difference.

Each has its pros and cons, some more suited to one specific goal, others
perhaps more appropriate for different kinds of test, such as the
optimization of parameters within nested models \cite{Esteves:2021}. Our
aim in this {\it Letter} is more straightforward than that. We shall
examine which strategy is likely to result in robust model selection
by the end of a nominal 20-year baseline.

\section{Redshift Drift}\label{redshiftdrift}
Objects moving with the Hubble flow display a redshift that depends on
distance and, if the expansion rate is evolving, on time as well. Thus,
measuring the change in redshift as the Universe expands constitutes
a direct, non-geometric probe of its dynamics, independent of any
assumptions concerning gravity and clustering \cite{Sandage:1962}.
An observed `redshift drift' may be readily interpreted as long as
the spacetime metric is consistent with the Cosmological principle,
which asserts that the Universe is homogeneous and isotropic (at
least on large scales).

Our principal concern in this {\it Letter} is how the time-dependent
redshift of a source may be used to infer the Universe's expansion
dynamics \cite{Corasaniti:2007,Quercellini:2012,Martins:2016}. The 
required measurements will be made with various instruments, 
including the ELT High-Resolution Spectrograph in the approximate 
redshift range $2\lesssim z\lesssim 5$ \cite{Liske:2008,Liske:2014}.

The first (and second) time derivatives of the redshift have been
derived by many workers (e.g., ref.~\cite{Weinberg:1972,Liske:2008,Martins:2016}),
so we shall simply quote their key results. Since the cosmological redshift,
$z$, between the time of emission, $t_e$, when the expansion factor
was $a(t_e)\equiv a_e$, and the observation time, $t_0$, when $a(t_0)\equiv
a_0$, is given as
\begin{equation}
1+z={a_0\over a_e}\;,\label{eq:z}
\end{equation}
its derivative in terms of $t_0$ is simply
\begin{equation}
{dz\over dt_0} = [1+z(t_0)]H(t_0)-{a_0\over a_e^2}{da_e\over dt_e}
{dt_e\over dt_0}\;,\label{eq:dzdt0}
\end{equation}
where the Hubble parameter is defined to be
\begin{equation}
H(t) \equiv {1\over a(t)}{da(t)\over dt}\;.\label{eq:H}
\end{equation}
We also have
\begin{equation}
dt_0=[1+z(t_0)]\,dt_e\;,\label{eq:dt0}
\end{equation}
and so
\begin{equation}
{dz\over dt_0}=[1+z(t_0)]H_0-H(z)\;,\label{eq:dzdt0final}
\end{equation}
where $H_0\equiv H(t_0)$ is the Hubble constant today.

The redshift-dependent Hubble variable is often written in the form
\begin{equation}
H(z) = H_0\,E(z)\;,\label{eq:Hz}
\end{equation}
where the dimensionless function
\begin{eqnarray}
E^2(z)&=&\Omega_m(1+z)^3+\Omega_r(1+z)^4+\nonumber\\
&\null&\qquad\Omega_{de}(1+z)^{3(1+w_{de})}+
\Omega_k(1+z)^2\;,\quad\label{eq:E}
\end{eqnarray}
explicitly shows the parametrization for any given cosmological model.

In the case of $\Lambda$CDM, $\Omega_m$, $\Omega_r$ and $\Omega_{de}$ are the
ratios of energy density for, respectively, matter ($\rho_m$), radiation ($\rho_r$)
and dark energy ($\rho_{de}$) over the critical density $\rho_c\equiv 3c^2H_0^2/8\pi G$.
In addition, $w_{de}$ is the equation of state parameter for dark-energy,
defined as $w_{de}\equiv p_{de}/\rho_{de}$, where $p$ is the pressure. The
ratio $\Omega_k\equiv -kc^2/(a_0^2\,\rho_c)$, in terms of the spatial
curvature constant $k$, is non-zero if the Universe is not spatially flat, i.e.,
if $k\not=0$.

The situation for $R_{\rm h}=ct$ is much simpler because $a(t)=t/t_0$
\cite{Melia:2007,MeliaShevchuk:2012,Melia:2016a,Melia:2016b,Melia:2020}.
In this case we have
\begin{equation}
H(t) \equiv {\dot{a}\over a}={1\over t}\label{eq:HRh}
\end{equation}
and, since $(1+z)=t_0/t_e$ from Equation~(\ref{eq:z}), we also find that
$(1+z)=H(t_e)/H(t_0)$, or
\begin{equation}
H(t_e)=H_0[1+z]\;.\label{eq:HRh2}
\end{equation}
Thus, it is easy to see from Equation~(\ref{eq:dzdt0final}) that
\begin{equation}
{dz\over dt_0}=0\label{eq:dzdt0Rh}
\end{equation}
at all redshifts for this model.

The surveys do not measure $\Delta z$ directly, but instead monitor
the spectroscopic velocity shift $\Delta v$ associated with $\Delta z$
over a time interval $\Delta t$, where
\begin{equation}
\Delta v={c\Delta z\over 1+z}={c\Delta t\over 1+z}{dz\over dt_0}\;.\label{eq:Dv}
\end{equation}
For the model selection we seek in this {\it Letter}, we shall therefore
be comparing the drift rates
\begin{equation}
{\Delta v\over \Delta t}(\Lambda{\rm CDM})=c H_0\left[1-{E(z)\over 1+z}\right]\label{eq:DvDt1}
\end{equation}
for {\it Planck}-$\Lambda$CDM, with
\begin{equation}
{\Delta v\over \Delta t}(R_{\rm h}=ct)=0\label{eq:DvDt2}
\end{equation}
for $R_{\rm h}=ct$.

\begin{figure}
\centering
\includegraphics[width=3.3in]{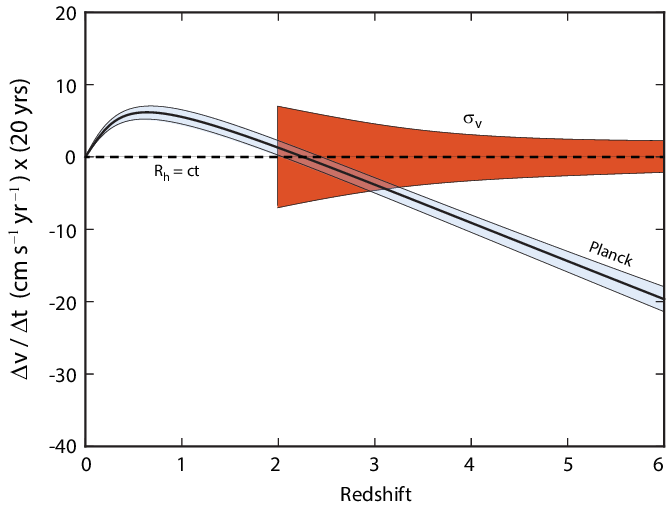}
\caption{Spectroscopic velocity shift $\Delta v/\Delta t$ predicted by 
{\it Planck}-$\Lambda$CDM (solid black) along with the $3\sigma$ C.L. (teal, bounded by thin black 
lines) after 20 years of monitoring with the ELT HIRES (Liske et al. 2008). The dashed black line
represents the zero redshift drift expected in the $R_{\rm h}=ct$ universe. The red shaded region
shows the expected $1\sigma$ error for this baseline as a function of redshift. For example, after 
20 years of monitoring, the ELT measurements should distinguish between {\it Planck}-$\Lambda$CDM 
and $R_{\rm h}=ct$ at better than $3\sigma$ for $z\gtrsim 3.6$.}
\label{fig1}
\end{figure}

The baseline required to discern the expected difference between
$\Lambda$CDM and $R_{\rm h}=ct$ depends on the measurement uncertainty
$\sigma_{v}$. According to \cite{Liske:2008}, the ELT-HIRES is
expected to reach an accuracy
\begin{equation}
\sigma_{v} = 1.35 {2370\over {\rm S/N}}\sqrt{{30\over N_{\rm QSO}}}\left({5
\over 1+z_{\rm QSO}}\right)^\alpha\;{\rm cm}\;{\rm s}^{-1}\;,\label{eq:sigmaELT}
\end{equation}
where $\alpha=1.7$ for $z\le 4$, and $\alpha=0.9$ for $z>4$, with a ${\rm S/N}$
of approximately 1,500 after 5 years of monitoring $N_{\rm QSO}=10$ quasars.

The Cosmic Accelerometer \cite{Eikenberry:2019} is a low-cost version of
the ELT-HIRES experiment, relying on the acquisition of off the shelf-equipment.
No detailed feasibility study has yet been carried out, though a spectroscopic
velocity uncertainty was quoted in the proposal white paper as part of the
Astro2020 decadal survey:
\begin{equation}
\sigma_{v}=\sigma_c\left({6\over t_{\rm exp}}\right)^{1/2}
\;{\rm cm}\;{\rm s}^{-1}\;,\label{eq:sigmaCA}
\end{equation}
where $\sigma_c=1.5$ and $t_{\rm exp}$ is the baseline in years. The expected
redshift range is similar to that of ELT, i.e., $2\lesssim z\lesssim 5$.

The Acceleration Programme \cite{Cooke:2020} will use ELT as described above,
also characterized by the uncertainty in Equation~(\ref{eq:sigmaELT}), though
it will attempt to measure the difference in drift between two non-zero
redshifts along the same line-of-sight:
\begin{equation}
{\Delta v_{12}\over \Delta t}= cH_0\left[{E(z_2)\over {1+z_2}}-{E(z_1)\over
{1+z_1}}\right]\;,\label{eq:diff}
\end{equation}
where $z_2$ and $z_1$ are, respectively, the redshift of the reference and
intervening sources, with $z_1<z_2$. For some choices of redshifts, the
drift signal can be larger than the case with $z_1=0$.

\section{Results}
We adopt concordance $\Lambda$CDM model parameters \cite{PlanckVI:2020}, 
$k=0$, $\Omega_m=0.3153\pm0.0073$, $\Omega_l=0.6847\pm0.0073$, $H_0=67.36\pm0.54$ km 
s$^{-1}$ Mpc$^{-1}$ and $w_{\rm de}=-1$, but also show the redshift drift predicted by 
this model within $\pm3\sigma$ of the optimal value via error propagation from  
the {\it Planck} uncertainties. $\Delta v$ for the standard model after 20 years of 
monitoring is shown as a function of $z$ (solid black curve) in figure~\ref{fig1}. The 
teal shaded region corresponds to the $3\sigma$ C.L. This $3\sigma$ uncertainty is 
typically $\sim 2.4\%$ over the range of redshifts sampled here. By comparison, the 
thick dashed line shows the strictly zero redshift drift expected in $R_{\rm h}=ct$.

The red shaded region in this figure shows the crucial $1\sigma$ uncertainty
in the spectroscopic velocity measurement (Eq.~\ref{eq:sigmaELT}) after 20 years
of monitoring, with a truncation at $z=2$, below which the Ly-$\alpha$ forest
falls in the unmeasurable ultraviolet portion of the spectrum. Clearly, the factor
guiding the confidence with which one of these models is selected by the redshift
drift data will be the uncertainty in the ELT Hires measurements, rather than the
measurement errors in {\it Planck} which, at this stage, are much smaller. We thus 
see that, with the ELT-HIRES experiment, one of these two models should be favored 
over the other at a confidence level of $3\sigma$ for sources at $z\gtrsim 3.6$.

\begin{figure}
\centering
\includegraphics[width=3.3in]{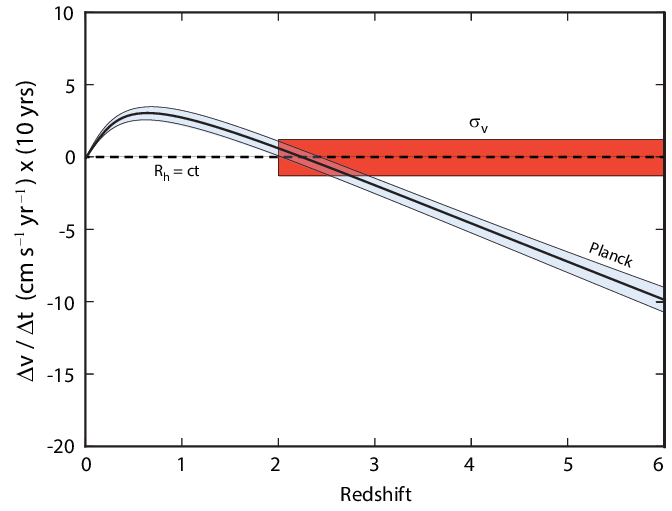}
\caption{Similar to Figure~\ref{fig1}, except now with the red $1\sigma$ error region corresponding
to the Cosmic Accelerometer (Eikenberry et al. 2019), and the spectroscopic velocity shift
estimated after 10 years of monitoring. In this case, the Cosmic Accelerometer measurements should
distinguish between {\it Planck}-$\Lambda$CDM and $R_{\rm h}=ct$ at better than $3\sigma$
for $z\gtrsim 2.6$.}
\label{fig2}
\end{figure}

\begin{figure}
\centering
\includegraphics[width=3.3in]{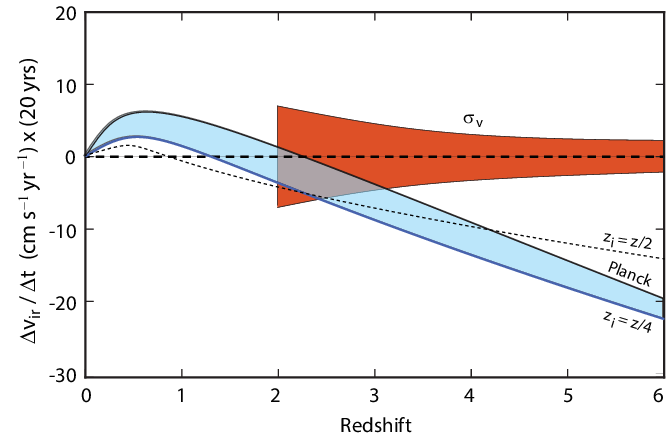}
\caption{Similar to Figure~\ref{fig1}, except now showing the prediction of {\it Planck}-$\Lambda$CDM
(solid thin black) and the differential redshift drift proposed for the Acceleration Programme
(Cooke 2020), with intermediate redshifts $z_i=z/4$ (solid thick blue)
and $z_i=z/2$ (dotted thin black). The $1\sigma$ error region is the same as that in Figure~\ref{fig1}.}
\label{fig3}
\end{figure}

The spectroscopic velocity uncertainty (Eq.~\ref{eq:sigmaCA}) is smaller than that of
the ELT-HIRES (Eq.~\ref{eq:sigmaELT}). In figure~\ref{fig2}, we therefore show the
cumulative velocity shift for $\Lambda$CDM and $R_{\rm h}=ct$ as a function of redshift
for a baseline of 10 years. The curves have the same meaning as those in figure~\ref{fig1},
though the red shaded region is now smaller, allowing the redshift drift measurements
to favor one of these models over the other by the same confidence level (i.e.,
$3\sigma$), in half the monitoring time. In addition, it might suffice to sample
sources even closer to us, i.e., at $z\gtrsim 2.6$, rather than $\gtrsim 3.6$.

Finally, we show in figure~\ref{fig3} the predicted differential drift as a function
of redshift (which is now $z_2$) expected over a 20-year baseline for two intermediate
redshifts, $z_1=z_2/4$ and $z_1=z_2/2$. The teal shaded region highlights the difference
between the redshift drift in the original ELT experiment (as in fig.~\ref{fig1}), and
the differential value corresponding to $z_1=z_2/4$. As one can see, this approach should
produce results favoring one model over the other at a confindence level of $3\sigma$
with sources as close as $z\sim 3$ instead of $\sim 3.6$, though with the same baseline
of 20 years.

\section{Discussion}\label{discussion}
Though our principal focus in this {\it Letter} has been a direct head-to-head comparison
between $R_{\rm h}=ct$ and Planck-$\Lambda$CDM, several other cosmological models have been
proposed in recent years to address the growing tension between the standard model and the
ever improving observations. 

\begin{figure}
\centering
\includegraphics[width=3.3in]{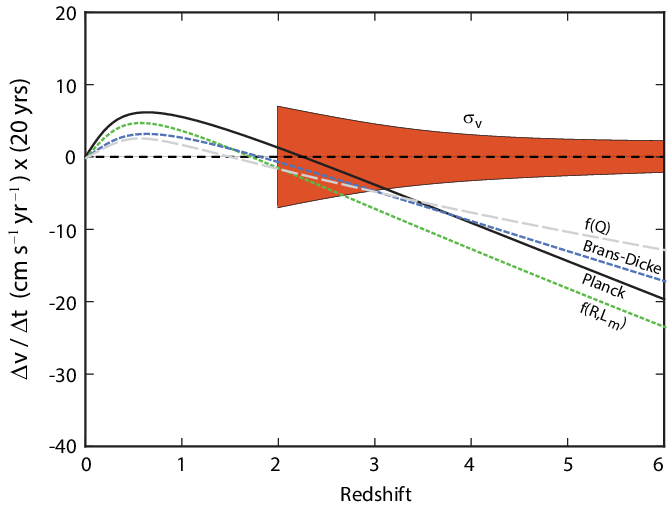}
\caption{Spectroscopic velocity shift $\Delta v/\Delta t$ predicted by {\it Planck}-$\Lambda$CDM
(solid black) and four other models, after 20 years of monitoring with the ELT HIRES: (i) 
$R_{\rm h}=ct$ {\sl flat dashed black}; (ii) a typical $f(R,L_m)$ model {\sl short dashed green}; 
(iii) the Brans-Dicke theory {\sl dashed blue}; and (iv) a typical $f(Q)$ model {\sl long 
dashed grey}. The red shaded region represents the expected $1\sigma$ error for this baseline 
as a function of redshift.}
\label{fig4}
\end{figure}

\begin{figure}
\centering
\includegraphics[width=3.3in]{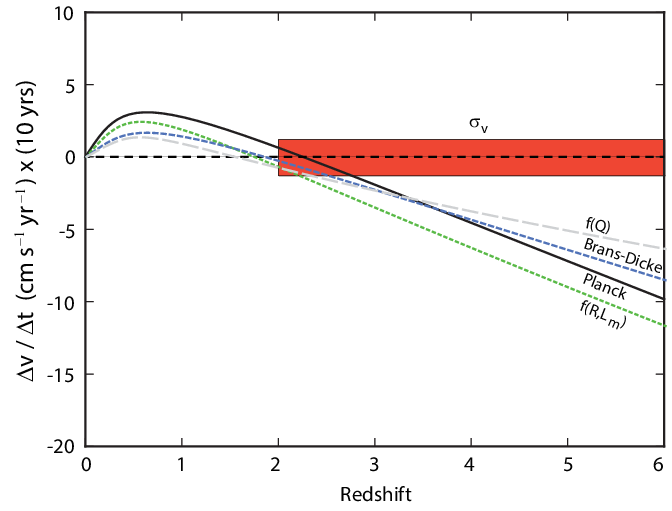}
\caption{Similar to Figure~\ref{fig4}, except now with the $1\sigma$ error region corresponding
to the Cosmic Accelerometer, and the spectroscopic velocity shift estimated after 10 years of
monitoring. In this case, the Cosmic Accelerometer measurements should distinguish between
$R_{\rm h}=ct$ and all the other models at better than $3\sigma$ for $z\gtrsim 4$.}
\label{fig5}
\end{figure}

In figures~\ref{fig4} and \ref{fig5}, we summarize these results analogously to those
shown in figures~\ref{fig1} and \ref{fig2} for three additional representative cosmologies,
which may be described as follows. The so-called $f(R,L_m)$ models constitute various
modified gravity theories in which dark energy is understood as an effective geometric
quantity, with the added flexibility that the Ricci curvature, $R$, may be coupled to
the Lagrangian density of matter, $L_m$ \cite{Bertolami:2007,Harko:2010}. To study their
cosmological implications, a suitable parametrization is adopted, such that
\begin{equation}
H(z) = H_0\left[A(1+z)^3+B+\epsilon\ln(1+z)\right]^{1/2}\;,\label{eq:Lm}
\end{equation}
where $H_0$, $A$, $B$ and $\epsilon$ are free parameters \cite{Myrzakulova:2023}. Their
optimization using cosmic chronometers and the Pantheon SN sample yields the values:
$H_0=68.0^{+1.5}_{-1.6}$ km s$^{-1}$ Mpc$^{-1}$, $A=0.34^{+0.11}_{-0.11}$, $B=1-A$,
and $\epsilon=-0.16^{+0.80}_{-0.79}$. The redshift drift predicted by this class of
models is shown as a short dashed green curve in figures~\ref{fig4} and \ref{fig5}.

An alternative approach to modifying general relativity, known as $f(Q)$ theory, attempts to
eliminate the dark sector by extending Einstein's equations into the quantum domain. An
illustrative choice of equation-of-state for the stress-energy tensor yields the following
Hubble parameter \cite{Koussour:2023},
\begin{equation}
H(z) = H_0\left[(1-\gamma)(1+z)^3+\gamma\right]^{1/2n}\;,\label{eq:fQ}
\end{equation}
in which the free parameters $H_0$, $\gamma$, and $n$, may be optimized using various
selected data. For example, fitting this function with cosmic chronometer $H(z)$ measurements,
one finds $H_0=65.5^{+4.3}_{-4.6}$ km s$^{-1}$ Mpc$^{-1}$, $\beta=0.33^{+0.39}_{-0.24}$
and $n=1.16^{+0.14}_{-0.15}$, where $\gamma\equiv (1+3\beta)^{-1}$. The redshift drift
predicted by this model is shown as a long dashed grey curve in figures~\ref{fig4} and
\ref{fig5}.

Finally, we consider a model in which dark energy is handled in the framework of generalized
Brans-Dicke theory with a self-interacting potential \cite{Tripathy:2023}. The Hubble
parameter in this scenario may be parametrized according to
\begin{equation}
H(z) = \delta + \beta(1+z)^\gamma\;.\label{eq:BD}
\end{equation}
An optimization of this expression using cosmic chronometer data yields the parameter
values: $H_0=64.57\pm23.53$ km s$^{-1}$ Mpc$^{-1}$, $\delta=36.67\pm9.58$ km s$^{-1}$ Mpc$^{-1}$,
$\beta=27.90\pm7.09$, and $\gamma=1.58\pm0.17$. The redshift drift predicted by this
type of model is shown as a dashed blue line in figures~\ref{fig4} and \ref{fig5}.

We draw several conclusions from this comparative study. First, 20 years of monitoring with
ELT may not be able to distinguish between Planck-$\Lambda$CDM and the $f(R,L_m)$, $f(Q)$
and Brans-Dicke based models for sources at $z\lesssim 5$. The Cosmic Accelerometer approach
may be able to discern $1-2\sigma$ differences between some of these models over the same
baseline, but only for similarly distant sources.

The $R_{\rm h}$ cosmology stands out in this group because it predicts strictly zero
redshift drift at all redshifts, while the rest of the models all need to account for
various phases of decelerated and accelerated expansion. As such, 10 years of Cosmic
Accelerator monitoring should be able to distinguish it from the rest of the models
at a confidence level of $3\sigma$ for sources at $z\gtrsim 4$. It should attain the
same level of confidence for sources at $z\lesssim 3.5$ after 20 years of monitoring.

\section{Conclusion}\label{conclusion}
The comparative study proposed in this {\it Letter} is motivated
by the growing observational evidence favoring the $R_{\rm h}=ct$
cosmology over the current standard model. Interest in implementing
the innovative idea of actually measuring the cosmic expansion in
real time has been growing following feasibility studies suggesting
that the next generation of telescopes, such as ELT, may have the
capability of assessing the redshift drift after a monitoring
campaign lasting one to two decades.

Several other projects are also aiming to measure the redshift
drift of selected quasar sources, so the prospects of carrying out the model
selection we have discussed in this {\it Letter} extend beyond the direct
comparison we have just described. For example, ESPRESSO is a relatively
new high-resolution spectrograph on ESO’s Very Large Telescope (VLT). It 
was designed for ultra-high radial-velocity precision and extreme spectral 
fidelity \cite{Pepe:2021}. The challenge is always to achieve the highest 
possible radial-velocity resolution, and ESPRESSO is expected to achieve
a precision of $\sim 10$ cm s$^{-1}$ with a baseline of a decade or more.
As one can see from figures~1-5, ESPRESSO may therefore be able assist
in model selection between $R_{\rm h}=ct$ and other cosmologies that
accelerate, but perhaps only with quasars at somewhat higher redshifts
and with an extended monitoring program, certainly at least several
decades long.

At radio wavelengths, FAST will observe damped Lyman-$\alpha$ absorbers in HI 
21 cm systems \cite{Lu:2022}, covering most of the Northern hemisphere at low 
redshifts ($0-0.35$), thereby complementing the planned ground-based Lyman-$\alpha$ 
forest projects, such as ELT, that will focus more on the Southern hemisphere 
and at higher redshifts ($1.5-5$). Its goal will be to determine the cosmic
expansion rate by measuring the redshift evolution of these HI 21 cm absorption-line 
systems over a decade or longer time span, on par with the other monitoring
programs we have been discussing \cite{Kang:2024}.

A similar approach will be followed by SKA, whose goal is to measure the spectral 
drift in the neutral hydrogen (HI) emission signals of galaxies to estimate their 
redshift drift in the redshift range of $0-1$. Unlike FAST, however, SKA will
observe predominantly the Southern hemisphere, thus again complementing the 
other monitoring campaigns. If SKA performs optimally and detects a billion 
galaxies, its measurement of redshift drift could achieve a precision of one 
percent \cite{Kloeckner:2015}.

If the upcoming measurement of redshift drift is strictly zero, this would 
completely invalidate the various phases of deceleration and acceleration
predicted by $\Lambda$CDM, along with many other categories of models, and 
redirect attention to the zero active mass condition (i.e., $\rho+3p$) 
underlying the $R_{\rm h}=ct$ cosmology. The consequences would be substantial. 
For example, it would obviate the need for inflation.

Our analysis has shown that this potentially critical outcome may be reached in 
only 10 years, using the Cosmic Accelerometer approach, if the survey functions 
as projected. A simple yes/no answer is all we shall need.

\acknowledgments I am grateful to the anonymous referees for their
helpful comments that have led to an improved presentation of the material in
this manuscript.

\bibliographystyle{eplbib}
\bibliography{ms}

\vskip0.1in
\section{Appendix}\label{Appendix}
We have been developing the $R_{\rm h}=ct$ cosmology for almost two
decades \cite{Melia:2003,MeliaShevchuk:2012,Melia:2020}. During this
time, more than 30 different kinds of cosmological data have been used 
in studies comparing its predictions with those of $\Lambda$CDM, at low
and high redshifts, based on a broad array of integrated and differential 
measures. These tests have consistently shown that $R_{\rm h}=ct$ accounts 
for the observations better than $\Lambda$CDM. A complete, more pedagogical 
description of this model, starting with its foundational underpinnings, and 
demonstrating its viability with the data, may be found in a recently published
monograph \cite{Melia:2020}.

This model was originally motivated by the discovery of a very awkward 
`coincidence' in the data, showing that the Universes's gravitational horizon 
is equal to the light-travel comoving distance since the Big Bang 
\cite{Melia:2003,Melia:2007,Melia:2018b}. This equality is possible only 
once in the history of the Universe, and it is happening right now 
\cite{MeliaShevchuk:2012}. The probability for such a coincidence is 
effectively zero. 

The most straightfoward `solution' to this perplexing outcome is that the 
equality $R_{\rm h}=ct$ (giving rise to the model's eponymous name) is true 
at all times, not merely at $t_0$. In that case, the probability of measuring
this equality would always be one.

The cosmic expansion rate implied by this equality via the Friedmann 
equations is constant, however, with an expansion factor $a(t)=t/t_0$, 
which contrasts noticeably with the variable expansion rate predicted 
in the standard model. Not surprisingly, therefore, early work with 
this hypothesis largely involved the acquisition of empirical evidence 
supporting such an unexpected scenario, in the face of significant
skepticism derived from $\Lambda$CDM's general degree of success
accounting for many broad features in the data. Yet test after test 
(now numbering over 30) throughout the redshift history of the Universe,
have all indicated that the observations quite compellingly
favor $R_{\rm h}=ct$ over the standard model. The sample of
such tests listed in Table~2 of ref.~\cite{Melia:2018e} suggests
that the `score' is at least 30 to 0 in favor of the former 
model.

More recent observations by JWST create an even bigger problem for
the standard model because they conflict with the timeline it predicts
for the formation of large-scale structure
\cite{Harikane:2022,Donnan:2022,Naidu:2022a,Bradley:2022}.
But they agree almost exactly with the evolution predicted by 
$R_{\rm h}=ct$ \cite{Melia:2023b,Melia:2023c}.

A deeper probe of its fundamental origin and viability has shown
that $R_{\rm h}=ct$ is really an updated version of $\Lambda$CDM, 
with one crucial additional constraint: the zero active mass condition 
from general relativity, i.e., $\rho+3p=0$, in terms of the total energy 
density $\rho$ and pressure $p$ in the cosmic fluid. One may see this 
directly from the Friedmann equations, which reduce to $a(t)=t/t_0$ when 
$p=-\rho/3$. 

More recent theoretical work shows that the zero active mass
condition may be necessary for the correct usage of FLRW in a cosmic 
setting \cite{Melia:2022b,Melia:2023}. The choice of lapse function, 
$g_{tt}=1$, in the FLRW spacetime metric precludes any acceleration, 
explaining why the zero active mass equation-of-state is necessary to 
produce an expansion with $a(t)=(t/t_0)$ at all redshifts.

This appears to be true even in the early Universe, which provides
straightforward solutions to many conflicts and hurdles in the standard
model. For example, $R_{\rm h}=ct$ completely avoids the monopole
problem \cite{Melia:2023f}, the cosmic entropy anomaly \cite{Melia:2021d}
and the trans-Planckian anomaly \cite{Melia:2020b}. 

On a broader scale, this model impacts every area in which $\Lambda$CDM 
has a major problem or inconsistency. The $R_{\rm h}=ct$ universe completely 
eliminates the CMB temperature \cite{Melia:2013c} and electroweak 
\cite{Melia:2018a} horizon problems, without needing inflation. It also
provides a natural explanation for the cutoff $k_{\rm min}$ observed
in the primordial power spectrum \cite{Melia:2019b}. 

In this {\it Letter}, we have described one of the few remaining types 
of test that may be used to clearly distinguish between $R_{\rm h}=ct$
and the standard model. Indeed, a determination of whether or not the 
Universe exhibits redshift drift is arguably more important and
compelling than all the other tests we have described in this Appendix.

\end{document}